\documentclass[11pt,a4paper,english,nofootinbib,,superscriptaddress]{revtex4}
\usepackage{lmodern}

\usepackage[T1]{fontenc}
\usepackage[latin9]{inputenc}
\usepackage{babel}

\usepackage{amsmath}
\usepackage{graphicx}
\usepackage{amssymb}
\usepackage{esint}
\usepackage[unicode=true, pdfusetitle,
 bookmarks=true,bookmarksnumbered=false,bookmarksopen=false,
 breaklinks=false,pdfborder={0 0 1},backref=false,colorlinks=false]
 {hyperref}
\def\b{\begin{equation}}
\def\e{\end{equation}}

\makeatletter
\@ifundefined{textcolor}{}
{%
 \definecolor{BLACK}{gray}{0}
 \definecolor{WHITE}{gray}{1}
 \definecolor{RED}{rgb}{1,0,0}
 \definecolor{GREEN}{rgb}{0,1,0}
 \definecolor{BLUE}{rgb}{0,0,1}
 \definecolor{CYAN}{cmyk}{1,0,0,0}
 \definecolor{MAGENTA}{cmyk}{0,1,0,0}
 \definecolor{YELLOW}{cmyk}{0,0,1,0}
 }

\usepackage{latexsym}\usepackage{bm}

\makeatother

\makeatother

\begin{document}

\title{Casimir Effect For a Scalar Field via Krein Quantization}

\author{H. Pejhan}
\email{h.pejhan@piau.ac.ir}
\affiliation{Department of Physics, Science and Research Branch, Islamic Azad University, Tehran, Iran}
\author{M.R. Tanhayi}
\email{m_tanhayi@iauctb.ac.ir}
\affiliation{Department of Physics, Islamic Azad University, Central Branch, Tehran, Iran}
\author{M.V. Takook}
\email{takook@razi.ac.ir}
\affiliation{Department of Physics, Science and Research Branch, Islamic Azad University, Tehran, Iran}

\date{\today}
\begin{abstract}

In this work, we present a rather simple method to study the Casimir effect on a spherical shell for a massless scalar field with Dirichlet boundary condition by applying the indefinite metric field (Krein) quantization technique. In this technique, the field operators are constructed from both negative and positive norm states. Having understood that negative norm states are un-physical, they are only used as a mathematical tool for renormalizing the theory and then one can get rid of them by imposing some proper physical conditions.

\end{abstract}
\maketitle

\section{Introduction}

The study of states with negative norm has a long story dating back to Dirac's work in 1942 \cite{Dirac}. Followed by Gupta and Bleuler in 1950, such states were used to remove the infrared (IR) divergence of QED \cite{Gupta}. In this way it is proved that quantization of the minimally coupled massless scalar field in de Sitter (dS) space can be done covariantly by the help of negative norm states \cite{Gazeau, De Bievre}. In other words, due to the famous `zero-mode' problem \cite{Allen, Gazeau}, one can not define a proper dS invariant vacuum state with only positive norm states, so a Gupta-Bleuler type construction based on Krein space structure is needed \cite{Gazeau}. This method provides a proposal to calculate graviton propagator in dS background in the linear approximation, without any pathological behaviour for largely separated points \cite{Behroozi}. [This result is in agreement with other works as long as it was shown that IR divergence of graviton propagator in one loop approximation do not appear in an effective way as a physical quantity because of gauge dependency \cite{AllenTuryn,Antoniadis,Higuchi}.]\\
Utilizing Krein space method, in the calculation of the expectation value of energy-momentum tensor, the infinite term does not appear, it means that the vacuum energy vanishes without any need of reordering the operators. Furthermore, in the interacting QFT this method works well in removing the singular behaviours of Green's function at short relative distances (UV divergence) except the light cone singularity \cite{Takook}. Recently, it was shown that the one-loop effective action for QED is regularized in a simple way in Krein space method \cite{Refaei}. The magnetic anomaly and Lamb shift are also studied in \cite{Forghan}.

In Ref. \cite{Khosravi}, through this approach, the Casimir force between the parallel plates in flat space has been calculated which simply gave the correct result. In this paper, we study this effect for a spherical shell with the Dirichlet boundary condition in which the oscillation modes are defined in Krein Space. It is shown that this method has some advantages over the usual methods in discussion of QFT.

The layout of the paper is as follows: The method of Krein space is briefly reviewed in section 2. In section 3, through this method, we calculate the Casimir force and as it is shown, negative norm states automatically regularize the theory and after renormalization the results are the same as the  previous related works. A brief discussion is done as a conclusion in section $4$. Finally, we have enclosed the paper with some details of mathematical calculations in two appendices.

\section{Krein Quantization: Basic Set-Up}

Massless minimally coupled scalar field in dS space plays an important role in the inflationary models and the linear quantum gravity in dS space. Allen proved \cite{Allen} that the covariant quantization of such a field in dS space cannot be constructed with only positive norm states. In Ref \cite{Gazeau}, a new version of indefinite metric field quantization or Krein space quantization method was used in order to quantize the massless minimally coupled scalar field in dS space covariantly, in which one should consider both negative and positive frequency solutions to preserve the causality and also eliminate the IR divergences.

Let us illustrate Krein quantization by giving a simple example. A free scalar field $\phi(\vec{x},t)$ which satisfies the Klein-Gordon equation
\begin{equation} \label{k} (\Box+m^2)\phi(x)=(\frac{\partial^2}{\partial t^2}-\nabla^2+m^2)\phi(\vec{x},t)=0,\end{equation}
has two sets of solutions:
\begin{equation} u_P(\vec{k},\vec{x},t)=\frac{e^{i \vec{k}\cdot\vec{x} -iwt}}{\sqrt{(2\pi)^{3}2w}} ,\;\,\mbox{and}\;\,\,u_N(\vec{k},\vec{x},t)=\frac{e^{-i\vec{k}\cdot\vec{x} +iwt}}{\sqrt{(2\pi)^{3}2w}}, \end{equation}
here the subscripts $P$ and $N$ are respectively referred to positive and negative frequency solutions and $w(\vec{k})=k^0=(\vec{k}\cdot\vec{k}+m^2)^{\frac{1}{2}}\geq 0 $. The inner product is defined by
\begin{equation} \label{in} (\phi_1,\phi_2)= i\int_{t=const}d^3x(\phi_1^\ast\partial_t\phi_2-\phi_2\partial_t\phi_1^\ast). \end{equation}

These modes are normalized by the following relations
\begin{equation}
\begin{aligned}
(u_P(\vec{k},\vec{x},t),u_P(\vec{k}',\vec{x},t))&=\delta(\vec{k}-\vec{k}'),\\
( u_N(\vec{k},\vec{x},t),u_N(\vec{k}',\vec{x},t))&=-\delta(\vec{k}-\vec{k}'),\\
( u_P(\vec{k},\vec{x},t),u_N(\vec{k}',\vec{x},t))&=0.
\end{aligned}
\end{equation}

Regarding the appearance of "un-physical states" in our method, one can circumvent the problem of propagating of such states and obtain physical results for measurable quantities by imposing following conditions on the quantum states and the probability amplitude (The "un-physical states" is used here in reference to the states with negative energies or frequencies\footnote{Noting that in usual indefinite metric quantizations which used in gauge QFT, the un-physical states have positive energy but due to the space-time metric signature, their norms become negative. However in our method, the un-physical states may have positive norms but their energies or frequencies are negative. This is the case of the presence of the spinor field in the theory such as QED, which the negative energy states have positive norms \cite{Forghan}.}):
\begin{itemize}
\item The physical boundary conditions have no effect on un-physical states. It means that in the Feynman diagrams such states only appear in the internal legs and in the disconnected parts of the diagrams \cite{Takook,Refaei}.
\item The unitarity of theory is survived by renormalizing the elements of the $S$ matrix (probability amplitude) as \cite{Forghan}
\begin{equation}\label{con2} S_{if}=\frac{\langle physical\; states, in|physical\; states, out\rangle}{\langle0 ,in|0, out\rangle}, \end{equation}
\end{itemize}
It is important to mention that, considering these conditions, in calculating the S-matrix elements or probability amplitudes for the physical states, negative energy states only appear in the internal line and in the disconnected parts of the Feynman diagrams. The un-physical states, which appear in the disconnected part of S-matrix elements, can be eliminated by renormalizing the probability amplitudes (the second condition). Presence of un-physical states in the internal line plays the key role in the renormalization procedure, so that it has been proved by considering quantum metric fluctuations, the divergences of Green's function in QFT do not appear any more \cite{M.V. Takook, Ford}.

Let us return to our example in which the field operator in Krein space is defined by \cite{Takook}
\begin{equation} \label{fk} \phi(t,\vec{x})=\frac{1}{\sqrt{2}}\Big(\phi_P(t,\vec{x})+\phi_N(t,\vec{x})\Big),\end{equation}
note that both $\phi_P$ and $\phi_N$ have been used and are defined by
\begin{equation}
\begin{aligned}
\phi_P(\vec{x},t)=&\int d^3 \vec{k}[a(\vec{k})u_P(\vec{k},\vec{x},t)+a^\dag(\vec{k})u_P^\ast(\vec{k},\vec{x},t)],\\
\phi_N(\vec{x},t)=&\int d^3 \vec{k}[b(\vec{k})u_N(\vec{k},\vec{x},t)+b^\dag(\vec{k})u_N^\ast(\vec{k},\vec{x},t)].
\end{aligned}
\end{equation}
The quantum theory is identified by defining the vacuum state as
$$a(\vec{k})|0\rangle=0,\,\,\,a^\dag(\vec{k})|0\rangle=|1_{k}\rangle,$$
\begin{equation} b(\vec{k})|0\rangle=0,\,\,\,b^\dag(\vec{k})|0\rangle=|\overline{1_{k}}\rangle,\end{equation}
where $|\overline{1_k}\rangle$ is an un-physical state. A significant difference with the standard QFT, which is based on the canonical commutation relation, lies in the requirement of the following commutation relations:

$$[a(\vec{k}),a^{\dag}(\vec{k}')]=\delta(\vec{k}-\vec{k}'),\;\; [a(\vec{k}),a(\vec{k}')]=[a^{\dag} (\vec{k}),a^{\dag}(\vec{k}')]=0,$$
$$[b(\vec{k}),b^{\dag}(\vec{k}')]= -\delta(\vec{k}-\vec{k}'),\;\;[b(\vec{k}),b(\vec{k}')]=[b^{\dag} (\vec{k}),b^{\dag}(\vec{k'})]=0,$$
\begin{equation}[a(\vec{k}),b(\vec{k})]=[a(\vec{k}),b^{\dag} (\vec{k})]=[a^{\dag} (\vec{k}),b(\vec{k})]=[a^{\dag} (\vec{k}),b^{\dag}(\vec{k})]=0.\end{equation}
In the next section, we will consider the field operator (\ref{fk}) in presence of the boundary condition.

\section{Considering the Boundary Condition: Casimir Effect}

An attractive force between two uncharged, parallel conducting plates was predicted by Casimir \cite{Casimir} hence this is called Casimir effect. This effect can be regarded as the physical manifestation of the zero-point energy and its quantum nature can be understood by the zero-point fluctuation of the quantum field which has to satisfy boundary conditions. A great deal of techniques have been proposed for studying the Casimir effect, e.g., direct mode summation \cite{Lambiase}, zeta function regularization \cite{Leseduarte}, Green's functions \cite{Milton} and stress-tensor methods \cite{Deutch}. For the sphere, in the literature a standard approach based on Boyer's work is usually followed in which instead of directly calculating the zero-point energy $E$ for a sphere, one calculates the difference $\Delta{E}$ in zero-point energy between two configurations\footnote{This method is similar to what was proposed by Casimir for two parallel conducting plates (see Ref. \cite{CasimirP}).}: the system is considered as a large spherical shell of radius $R$ enclosing the quantization universe and a small concentric sphere of variable radius \cite{Boyer,Nesterenko}. Then the zero-point energy of the inner sphere, ${E(a)}$, will be the charge in the zero-point energy of the total system when the radius of the inner sphere is changed from a radius $a$ to some radius $R/{\eta}$, $\eta > 1$. Namely,
$$ E(a)= \lim_{R\rightarrow\infty}\Delta{E(a,R)},$$
in which $\Delta{E(a,R)}\equiv (E_{in}+E_{out})-(E'_{in}+E'_{out})$, $E_{in}$ ($E_{out}$) and $E'_{in}$ ($E'_{out}$) are the inside (outside) zero-point energy of sphere with radius $a$ and $R/{\eta}$, respectively. Note that the value of $\lim_{R\rightarrow\infty}\Delta{E(a,R)}$ is independent of $R$ and the value chosen for $\eta$ provided that $R/{\eta}\gg a$ and $ R\gg R/{\eta}$. Choosing $\eta\sim 1$ or equivalently $R\sim R/\eta$, results in an additional energy owing to the attraction of the inner comparison sphere to the outer sphere of the universe. This will not occur if $\eta > 1$ is held fixed while $R$ increases \cite{Boyer}.\\
Here, we use the Krein space field quantization technique to calculate the zero-point energy for a massless scalar field with Dirichlet boundary condition on a spherical shell. As it is shown by the help of the negative norm states the interpretation will be more easier than the standard method. By imposing the physical boundary condition on the field operator (\ref{fk}), only the positive norm states are affected, in other words, the negative modes do not interact with the physical states or real physical world, thus they can not be affected by the physical boundary conditions. So, the field operator can be written as:
\begin{equation} \label{1} \phi(\vec{x},t)=\sum_L [a(k_L) u_P(k_L,\vec{x},t)+a^\dag(k_L)u_P^\ast (k_L,\vec{x},t)]+\int d^3 k [b(k)u_N (k,\vec x,t)+b^\dag (k) u^\ast _N (k,\vec x,t)], \end{equation}
here $k_L$ are the eigen-frequencies of the system under consideration. For the case of a spherical shell of radius $a$, regarding to field operator (\ref{1}), the total vacuum energy (Casimir energy) is obtained as follows:
\begin{equation}\label{sp} E_{(sphere)}^{Krein}={\sum_{l=0}^{\infty}{(l+\frac{1}{2})} \sum_{n=0}^{\infty}{\omega_{nl}}}-\frac{a^3}{3\pi}\int_{0}^{\infty}k^3dk, \end{equation}
where the eigen-frequencies $\omega_{nl}$ are determined by the spherical Bessel function $ j_l(\omega a)=0$, [$j_l(z)={{\sqrt{\frac{\pi}{2z}}}}J_{l+1/2}(z)$] with $l=0,1,2,...$.  The second term at the right hand side of Eq. (\ref{sp}) appears due to the un-physical states where we want to convert it to a comparable term with the first one. In fact as mentioned, physical boundary conditions do not affect the un-physical states, so with good accuracy, zero-point energy density due to the such modes can be thought as the energy density of physical states within a large spherical shell $(R\rightarrow \infty)$ concentric to the sphere of radius $a$. Therefore, the zero-point energy density of un-physical states can be written as
\begin{equation} \label{'3.16'}\int_{0}^{\infty}k^3dk \approx \frac{1}{L^{3}} \sum_{l=0}^\infty(l+\frac{1}{2})\sum_{n=0}^\infty{\frac{\pi}{L}}[(n+\frac{1}{2})+\frac{l+1}{2}],\;\;\; L\gg 1.\end{equation}
Note that, for sufficiently large values of $L$, the term under the second summation can be considered as the allowable eigen-frequencies inside a large spherical shell, with radius $L$, which are determined by $\lim_{{a}\rightarrow\infty} j_l(\bar{\omega}_{nl}\;a)=0$,\footnote{In Appendix A, it is shown that this interpretation will be more accurate with increasing the order of magnitude of $L$.} and hence, the right hand side would be the vacuum energy of physical states inside a large spherical shell which is divided by the sphere's volume. On the other hand, the uniform energy distribution, due to the large radius of spherical shell, allows us to consider (\ref{'3.16'}) as the asymptotic behavior of zero-point energy density of physical states inside the large spherical shell.

Now with this approximate form, Eq. (\ref{sp}) can be written as:
\begin{equation} \label{sp'} E_{(sphere)}^{Krein}={\sum_{l=0}^{\infty}{(l+\frac{1}{2})} \sum_{n=0}^{\infty}{\omega_{nl}}} -\sum_{l=0}^\infty(l+\frac{1}{2})\sum_{n=0}^\infty\frac{\pi}{{\cal{L}}}[(n+\frac{1}{2})+\frac{l+1}{2}],\;\;\; \mbox{ where ${\cal{L}}\equiv \frac{3{\pi} L^4}{a^3}$}, \end{equation}
${\cal{L}}$ has the length dimension and its order of magnitude is much more than that of $L$, thus $\frac{\pi}{{\cal{L}}}[(n+\frac{1}{2})+\frac{l+1}{2}]$ can be specified by $\bar\omega_{nl}$ with high precision. Therefore, Eq. (\ref{sp'}) turns to:
\begin{equation} \label{3.16} E_{(sphere)}^{Krein}=\sum_{l=0}^\infty(l+\frac{1}{2})\sum_{n=0}^\infty(\omega_{nl}-\bar\omega_{nl}), \end{equation}
noting that $\lim_{{a}\rightarrow\infty} j_l(\bar{\omega}_{nl}\;a)=0$. In continue, it is suitable to write $E_{(sphere)}^{Krein}=\sum_{l=0}^\infty  E_l^{Krein}$ by defining
\begin{equation} \label{E1} E_l^{Krein}\equiv(l+\frac{1}{2})\sum_{n=0}^\infty(\omega_{nl}-\bar\omega_{nl}).\end{equation}

At this stage, we use Cauchy's theorem to convert the summation (\ref{E1}) in to the integral form which reads as
\begin{equation} \label{cuchy} E_l^{Krein}=\frac{l+1/2}{2\pi i}\Big(\oint_C dz\;z\frac{d}{dz}\ln {\frac{f(z,a)}{f(z,a\rightarrow\infty)}}\Big),\end{equation}
where $f(z,a)$ is defined by\footnote{It should be noted that $J_\nu(z)$, $\nu>-1$, has no complex roots \cite{Abramowitz}. Therefore according to  Cauchy's integral theorem (Eq. (\ref{cuchy})), $f(z,a)=J_\nu(a\;\textbf{z}e^{i\theta})$ or $J_\nu(a\;\textbf{z}e^{i|\theta|})$ leads to same result  for $E_l^{Krein}$.}
\begin{equation} \label{f(z,a)} f(z,a)\equiv J_\nu(a\;\textbf{z}e^{i|\theta|}), \;\;\;\;\;\;\;\; \nu=l+1/2=1/2,3/2,...,\end{equation}
($"\textbf{z}"$ and $"\theta"$ are respectively the modulus and argument of $z$.) The contour of integration consists of two parts. The first part is  a counterclockwise semicircular $C_\Lambda$, centered at the origin in the right half-plane and having a radius $\Lambda$ large enough to enclose all  the poles of $f(z,a)$. The second part of the contour is the straight line from $(0,i\Lambda)$ to $(0,-i\Lambda)$ on the imaginary axis. Note that $z$  in (\ref{cuchy}) can be taken as $ \lim_{m\rightarrow0}\sqrt{z^2+m^2}$, therefore in the integral along the segment $(i\Lambda, -i\Lambda)$ we can integrate once by parts and the nonintegral terms being cancelled \cite{Nesterenko}. Therefore, Eq. (\ref{cuchy}) becomes
\begin{equation} \label{bypart} E_l^{Krein}=\frac{l+1/2}{\pi }\int_0^\infty dy\;\ln {\frac{f(i y,a)}{f(i y,a\rightarrow\infty)}}+\frac{l+1/2}{2\pi i}\int_{C_{\Lambda(-\frac{\pi}{2},\frac{\pi}{2})}} z\;d\ln {\frac{f(z,a)}{f(z,a\rightarrow\infty)}},\end{equation}
where we have (Appendix A)
\begin{equation} \label{f1} \frac{f(i y,a)}{f(i y,a\rightarrow\infty)}=\frac{J_\nu(i y a)}{\lim_{a\rightarrow\infty}J_\nu(i y a)}=\frac{I_\nu(y a)}{\lim_{a\rightarrow\infty}I_\nu(y a)}=\sqrt{2\pi ay}e^{-ay}I_\nu(y a),\end{equation}
therefore one obtains
\begin{equation} \label{new} E_l^{Krein}=\frac{l+1/2}{\pi }\int_0^\infty dy\;\ln {(\sqrt{2\pi ay}e^{-ay}I_\nu(y a))}+{\frac{l+1/2}{2\pi i}\int_{C_{\Lambda(-\frac{\pi}{2},\frac{\pi}{2})}} z\;d\ln {\frac{f(z,a)}{f(z,a\rightarrow\infty)}}},\end{equation}
Eq. (\ref{new}) after relatively simple calculations, can be written as
$$ E_l^{Krein}=\frac{l+1/2}{\pi}{\int_0^\infty}dy\ln({2ay}K_\nu(ay)I_\nu(ay))$$
\begin{equation} \label{asymp} +{\frac{l+1/2}{\pi}{\int_0^\infty}dy\ln\frac{\lim_{a\rightarrow\infty}K_\nu(ay)}{K_\nu(ay)}} +\frac{l+1/2}{2\pi i}\int_{C_{\Lambda(-\frac{\pi}{2},\frac{\pi}{2})}} z\;d\ln{\frac{f(z,a)}{f(z,a\rightarrow\infty)}}. \end{equation}
$I_\nu(z)$ and $K_\nu(z)$ are modified Bessel functions. In Appendix B, it is shown that the sum of two last terms in (\ref{asymp}) vanishes, then we get
\begin{equation}
\begin{aligned} \label{=0}
E_l^{Krein}&=\frac{l+1/2}{\pi}{\int_0^\infty}dy\ln({2ay}K_\nu(ay)I_\nu(ay)),\\
&=\frac{l+1/2}{a\pi}{\int_0^\infty}dy\ln({2y}K_\nu(y)I_\nu(y)).
\end{aligned}
\end{equation}
For large $\nu$ the above integral behaves as follows
\begin{equation} \label{behave} E_l^{Krein}\simeq \frac{1}{a} \Big( -\frac{\nu^2}{2}-\frac{1}{128}+\frac{35}{32768{\nu}^2}+{\cal{O}}{({\nu}^{-3})}\Big).\end{equation}
Substituting (\ref{behave}) in (\ref{3.16}), because of the first two terms, leads to divergence, but it can be handled easily by rewriting  (\ref{3.16}) as \cite{Nesterenko}
\begin{equation}
\begin{aligned} \label{3.26} E_{(sphere)}^{Krein}&={\sum_{l=0}^{\infty}\widetilde{E}_l^{Krein}}  -\frac{1}{2a}{\sum_{l=0}^{\infty}(l+1/2)^2}-\frac{1}{128a}{\sum_{l=0}^{\infty}(l+1/2)^0},\\
&={\sum_{l=0}^{\infty}\widetilde{E}_l^{Krein}} -\frac{1}{2a}\zeta(-2,1/2)-\frac{1}{128a}\zeta(0,1/2)={\sum_{l=0}^{\infty}\widetilde{E}_l^{Krein}},
\end{aligned}
\end{equation}
where
$$\widetilde{E}_l^{Krein}\equiv{E}_l^{Krein}+\frac{(l+1/2)^2}{2a}+\frac{1}{128a}\simeq\frac{1}{a}\frac{35}{32768{\nu}^2},$$
note that $\zeta(-2,1/2)=\zeta(0,1/2)=0$ \cite{Wittaker,Gradshteyn}. The last sum in (\ref{3.26}) obviously converges and by the means of numerical  integration, we obtain
\begin{equation}\label{result} E_{(sphere)}^{Krein}=\frac{1}{a}0.002819...., \end{equation}
which is the Casimir energy for a  spherical shell. The Casimir force can be obtained accordingly $F=-\frac{d}{da}E$, which is finite. In Ref. \cite{C.M. Bender}, with greater accuracy, this force was calculated by making use of Green's function technique.

\section{Conclusion}

The advent of divergence in expectation value of the energy-momentum tensor and consequently in zero-point energy causes fundamental problems in modern physics \cite{Rugh}. Basically, the standard method is to apply the renormalization schemes to overcome this problem. As it is well known, in non-gravitational physics, only the energy differences are measurable, so one can rescale -or renormalize- the zero-point energy by subtracting infinite energy from quantum vacuum energy \cite{Birrell}. In this process, the vanishing vacuum expectation value of the energy-momentum tensor in Minkowski space-time is considered as the physical "renormalization condition" \cite{Kay}. However, when the gravity is taken into account, the situation is different and this method is no longer satisfactory because the entire energy-momentum is supposed to be the source of gravitational field. Therefore, the rescaling method cannot be expected to be valid generally.

Actually as mentioned, in the covariant quantization of the minimally coupled scalar field in de Sitter space-time, one needs to preserve the negative frequency solutions, which this leads to a modification of QFT by reinterpreting its formalism. The presence of negative frequency solutions (un-physical states), during the quantizing process, provides a natural tool for eliminating the singularity in QFT. [Of course, the additional conditions are needed to impose on the quantum states and the probability amplitude as well, to preserve the unitarity of the theory.] Through this method, vacuum expectation value of the energy-momentum tensor automatically vanishes without need of energy rescaling. In this paper, we generalized this method to calculate the Casimir energy for a spherical shell in which the result is the same as the previous related works, though, the starting point of our approach is not coincide with the conventional view ($i.e.$ Boyer's method \cite{Boyer}).

We must emphasize the fact that although the both methods, energy rescaling and Krein methods, predict the similar results (at least in Minkowski space-time), but the perspective and procedure of these two methods are different. In the standard method, the finite results are acquired by a process named the renormalization, however, in the Krein method, infinities are eliminated and theories are automatically renormalized (or at least, are regularized). Thus, it may provide a way to address the zero-point energy divergence problem in the presence of gravitational field, where the energy rescaling would not be satisfactory \cite{Birrell}. This case will be considered explicitly in the forthcoming papers.

\vspace{.7cm}

\noindent {\bf{Acknowledgements:}} M.V. Takook would like to thank Prof. J. Iliopoulos for very useful discussions on the unitarity problem. We also thank Mostafa Tanhayi-Ahari for his early cooperation in this work. We are grateful to B. Forghan and E. Ariamand for their useful comments.

\begin{appendix}
\section{Some useful relations}

In this appendix we first review the Hurwitz $\zeta$-function briefly and then collect some useful relations which used in this paper.

\textbf{I)} The Hurwitz $\zeta$-function is formally defined by \cite{Wittaker, Gradshteyn}
\begin{equation} \label{zetadef} \zeta(z,\alpha)=\sum_{l=0}^{\infty}{\frac{1}{(\alpha+l)^z}}, \end{equation}
where $\zeta(z,1)$ is the Riemann zeta function. For non-positive integer $z (=-n)$, the Hurwitz $\zeta$-function and Bernoulli polynomials are  related as:
\begin{equation} \label{Bernoulli} \zeta(-n,\alpha)=-\frac{B_{n+1}(\alpha)}{n+1},\end{equation}
where the first few Bernoulli polynomials are:
\begin{equation} \label{fewBernoulli} B_0(\alpha)=1,\;\;\;B_1(\alpha)=\alpha-\frac{1}{2},\;\;\;B_2(\alpha)= \alpha^2-\alpha+\frac{1}{6},\;\;\;B_3(\alpha)=\alpha^3-\frac{3}{2}\alpha^2+\frac{1}{2}\alpha, ...\end{equation}

\textbf{II)} Modified Bessel functions, $I_\nu(z)$ and $K_\nu(z)$, are defined by \cite{Abramowitz}
\begin{equation} \label{00a1} J_{\nu}(iz)=i^{\nu}I_{\nu}(z),\;\;\;\;\; K_{\nu}(z)=(\pi/2)i^{{\nu}+1}H_{\nu}^{(1)}(iz),\end{equation}
where $J_{\nu}(z)$ and $H_{\nu}^{(1)}(z)$ are Bessel and Hankel functions of the first kind, respectively. Asymptotic forms for large arguments  become
\begin{equation}\label{jadid}J_{\nu}(z)= \sqrt{\frac{2}{\pi z}}\left\{P_\nu(z)\cos\left[z-\left(\nu+\frac{1}{2}\right)\frac{\pi}{2}\right] -Q_\nu(z)\sin\left[z-\left(\nu+\frac{1}{2}\right)\frac{\pi}{2}\right]\right\},\;\;\;\;\; |arg\; z|<\pi, \end{equation}
\begin{equation} \label{A.2} K_\nu(z)=\sqrt{\frac{\pi}{2z}}e^{-z}\Big({P_\nu(i z)+iQ_\nu(i z)}\Big),\;\;\;\;\; |arg\; z|<3\pi/2,\end{equation}
\begin{equation} \label{A.3} I_\nu(z)=\sqrt{\frac{1}{2\pi z}}e^{z}\Big({P_\nu(i z)-iQ_\nu(i z)}\Big),\;\;\;\;\; |arg\; z|<\pi/2,\end{equation}
where
$$ P_\nu(z)\sim 1-\frac{(\mu-1)(\mu-9)}{2!(8z)^2}+\frac{(\mu-1)(\mu-9)(\mu-25)(\mu-49)}{4!(8z)^4}-...\;,$$
$$ Q_\nu(z)\sim \frac{(\mu-1)}{1!(8z)}-\frac{(\mu-1)(\mu-9)(\mu-25)}{3!(8z)^3}+...,\;\;\;\;\;\; \mu=4\nu^2.$$
It is of some interest to consider the accuracy of the asymptotic form (\ref{jadid}), taking only the first term,
$$ J_{\nu}(z)\approx\sqrt{\frac{2}{\pi z}}\cos\left[z-\left(\nu+\frac{1}{2}\right)\frac{\pi}{2}\right].$$
Clearly, the condition for validity of above equation is that the sine term be negligible; that is
$$ 8z \gg 4{\nu}^2-1.$$
For $\nu> 1$ the asymptotic region may  be far out. [Note: $j_l(z)={{\sqrt{\frac{\pi}{2z}}}}J_{l+1/2}(z)$]

\section{Mathematical relations underling the Eq. (\ref{asymp})}

Let us define $g\Big(z(\equiv \textbf{z} e^{i\theta}),a\Big)$ as
\begin{equation} \label{z1} g(z,a)\equiv K_\nu(a\;\textbf{z} e^{i(|\theta+\pi/2|-\pi/2)}),\;\;\;\;\;\; \nu=l+1/2=1/2,3/2,...\end{equation}
then it follows
\begin{enumerate}\label{pro}
\item $ g(\textbf{z}e^{i\theta},a)=K_\nu(a\;\textbf{z}e^{i\theta})$, \hspace{3cm}for \,\,\,$ -\pi/2\leq \theta\leq\pi/2 $,
\item $ g(\textbf{z}e^{i\theta_+},a)=g(\textbf{z}e^{i\theta_-},a)$,  \hspace{2.7cm} where $\theta_\pm=-\pi/2\pm\theta' $,
\item $ g(\textbf{z}e^{i\theta},a)$ has no zeros\footnote{It has no zeros as long as $K_\nu(z)$, for any real number $\nu$, has no roots in the region $|arg\; z|\leq \frac{\pi}{2}$ \cite{Abramowitz}. On the other hand note that, if we choose $ -\pi/2\leq \theta_{+}\leq\pi/2 $, the second property imposes $ g(\textbf{z}e^{i\theta_{+}},a)=g(\textbf{z}e^{i\theta_-},a),$ where  $-3\pi/2\leq \theta_{-}\leq-\pi/2 $, again one gets back to third property. }, \hspace{2.9cm}for \,\,\,$ -3\pi/2\leq \theta\leq\pi/2. $
\end{enumerate}
Now according to the Cauchy's theorem, it is easy to verify that we have
\begin{equation} \label{b1} \oint_{C'}dz\;z\frac{d}{dz}\ln\Big(\frac{g(z,{a\rightarrow\infty})}{g(z,a)}\Big)=0,\end{equation}
$C'$ is a counterclockwise contour of integration consists of the real axis $[R,-R]$ and a semicircle $C'_R$, in the down half-plane with radius $R$ which we let $R$ becomes infinitely large. After separating the contributions of different parts of the counter $C'$ and using integration by parts  (see Eq. (\ref{bypart})), we can write (\ref{b1}) as
\begin{equation} \label{b3} 2\int_{0}^{\infty} dx\;\ln\Big(\frac{\lim_{a\rightarrow\infty}K_\nu(x\;a)}{K_\nu(x\;a)}\Big)+\int_{C'_{R(-\pi,0)}}z\; d\ln\Big(\frac{g(z,{a\rightarrow\infty})}{g(z,a)}\Big)=0.\end{equation}
From (\ref{b3}), the last two terms in Eq. (\ref{asymp}) turns to
\begin{equation} \label{b4} \frac{l+1/2}{2\pi}\int_{\;C'_{R(-\pi,0)}}z\; d\ln\Big(\frac{g(z,a)}{g(z,{a\rightarrow\infty})}\Big)+\frac{l+1/2}{2\pi i}\int_{\;C_{\Lambda(-\frac{\pi}{2},\frac{\pi}{2})}} z\;d\ln{\frac{f(z,a)}{f(z,a\rightarrow\infty)}}. \end{equation}
Substituting $z\rightarrow{iz}$ in the last term, results in
\begin{equation} \label{b5} \frac{l+1/2}{2\pi}\int_{\;C'_{R(-\pi,0)}}z\; d\ln\Big(\frac{g(z,a)}{g(z,{a\rightarrow\infty})}\Big) +\frac{l+1/2}{2\pi}\int_{\;C_{\Lambda(-\pi,0)}} z\;d\ln{\frac{f(i z,a)}{f(i z,a\rightarrow\infty)}}, \end{equation}
in which $f(i z,a)$, is defined as (\ref{f(z,a)} and \ref{00a1})
\begin{equation} \label{f(iz,a)} f(i z,a)=J_\nu(a\;\textbf{z}e^{i|\theta+\pi/2|}) =J_\nu(a\;i\textbf{z}e^{i(|\theta+\pi/2|-\pi/2)})=i^\nu I_\nu(a\;\textbf{z}e^{i(|\theta+\pi/2|-\pi/2)}).\end{equation}
Note that $f(iz,a)$ and $g(z,a)$ have the similar properties, therefore we can rewrite (\ref{b5}) as follows:
\begin{equation} \label{b6'} \frac{l+1/2}{\pi}\Big(\int_{\;C'_{R(-\pi/2,0)}} z\;d\ln{\frac{K_\nu(z\;a)}{\lim_{a\rightarrow\infty}K_\nu(z\;a)}}+\int_{\;C_{\Lambda(-\pi/2,0)}} z\;d\ln{\frac{I_\nu(z\;a)}{\lim_{a\rightarrow\infty}I_\nu(z\;a)}}\Big).\end{equation}
At the large radius of the contour and after making use of (\ref{A.2}), (\ref{A.3}), one gets
\begin{equation} \label{b6} (\ref{b6'})=\frac{l+1/2}{\pi}\int_{\;C_{\infty(-\pi/2,0)}} z\;d\ln(P_\nu ^2(i az)+Q_\nu ^2(i az)),\end{equation}
the term under the integral falls off at large $z$ faster than $\frac{1}{z^2}$, thus at this limit this term can be neglected.

\end{appendix}

\end{document}